\documentclass{article}

\usepackage{graphicx}
\usepackage{authblk}
\usepackage{url}

\usepackage[a4paper, total={5.5in, 8in}]{geometry}

\usepackage{tikz,tikz-qtree,tikz-qtree-compat}
\usetikzlibrary{fit,topaths,calc,shapes,decorations,chains,scopes,arrows,positioning}

\tikzstyle{path} = [->,double,rounded corners=.1cm]
\tikzstyle{data} = [draw, rectangle, rounded corners = .07cm, align=center, inner sep = .2cm, outer sep = .1 cm]

\colorlet{clr_outofDB}{red!80!black}
\colorlet{clr_inDB}{green!50!black}

\newcommand{\drawtable}[5]
{
   \begin{scope}[shift={(-#1/2,-#2/2)}]
      \pgfmathsetmacro{\cellwidth}{#1/#3}
      \pgfmathsetmacro{\cellheight}{#2/#4}
      \draw[fill=#5!30] (0, #2) rectangle (#1, #2-\cellheight);
      \foreach \i in {1,...,#3}
         \draw[thin, #5!70] (\cellwidth*\i, 0) -- (\cellwidth*\i, #2);
      \foreach \i in {1,...,#4}
         \draw[thin, #5!70] (0, \cellheight*\i) -- (#1, \cellheight*\i);
      \draw[#5] (0, #2-\cellheight) -- (#1, #2-\cellheight);
      \draw[thick, #5] (0, 0) rectangle (#1, #2);
   \end{scope}
}

\newcommand{\drawcylinder}[4]
{
   \begin{scope}[shift={(0,#3/2)}]
      \draw[#4] (0, #2/2) ellipse (#1/2 and #3);
      \draw[#4] (#1/2, -#2/2) arc (0:-180:#1/2 and #3);
      \draw[#4] (-#1/2, #2/2) -- (-#1/2, -#2/2);
      \draw[#4] (#1/2, #2/2) -- (#1/2, -#2/2);
   \end{scope}
}

\begin{document}

\title{Learning Models over Relational Data:\\ A Brief Tutorial\thanks{This project has received funding from the European Union's Horizon 2020 research and innovation programme under grant agreement No 682588.}}

\author{Maximilian Schleich$^1$, Dan Olteanu$^1$,
  Mahmoud Abo-Khamis$^2$, \\ Hung Q. Ngo$^2$, XuanLong Nguyen$^3$}

\date{%
    $^1$University of Oxford\\%
    $^2$relational\underline{AI}\\
    $^3$University of Michigan\\[2ex]%
    {\small \url{https://fdbresearch.github.io}\hspace{2em}
    \url{https://www.relational.ai}}
}

\maketitle    
\begin{abstract}
This tutorial overviews the state of the art in learning models over relational databases and 
makes the case for a first-principles approach that exploits recent developments in database research. 

The input to learning classification and regression models is a training dataset defined by feature extraction queries over relational databases. 
The mainstream approach to learning over relational data is to materialize the training dataset, export it out of the database, and then learn over it using a statistical package. This approach can be expensive as it requires the materialization of the training dataset. An alternative approach is to cast the machine learning problem as a database problem by transforming the data-intensive component of the learning task into a batch of aggregates over the feature extraction query and by computing this batch directly over the input database. 

The tutorial highlights a variety of techniques developed by the database theory and systems communities to improve the performance of the learning task. They rely on structural properties of the relational data and of the feature extraction query, including
algebraic (semi-ring), combinatorial (hypertree width), statistical (sampling), or geometric (distance) structure. 
They also rely on factorized computation, code specialization, query compilation, and parallelization. 

\end{abstract}

\section{The Next Big Opportunity}

Machine learning is emerging as general-purpose technology just as computing became general-purpose 70 years ago. A core ability of intelligence is the ability to predict, that is, to turn the information we have into the information we need. Over the last decade, significant progress has been made on improving the quality of prediction by techniques that identify relevant features and by decreasing the cost of prediction using more performant hardware.

According to a 2017 Kaggle survey on the state of data science and machine learning among 16,000 machine learning practitioners~\cite{kaggle-survey}, the majority of practical data science tasks involve relational data: in retail, 86\% of used data is relational; in insurance, it is 83\%; in marketing, it is 82\%; while in finance it is 77\%. This is not surprising. The relational model is the jewel in the data management crown. It is one of the most successful Computer Science stories. Since its inception in 1969, it has seen a massive adoption in practice. Relational data benefit from the investment of many human hours for curation and normalization and are rich with knowledge of the underlying domain modelled using database constraints. 

Yet the current state of affairs in building predictive models over relational data largely ignores the structure and rich semantics readily available in relational databases. Current machine learning technology throws away this relational structure and works on one large training dataset that is constructed separately using queries over relational databases.

This tutorial overviews on-going efforts by the database theory and systems community to address the challenge of efficiently learning machine learning models over relational databases. It invariably only highlights some of the representative contributions towards this challenge, with an emphasis on recent contributions by the authors.
The tutorial does not cover the wealth of approaches that use arrays of GPUs or compute farms for efficient machine learning. It instead puts forward the insight that an array of known and novel database optimization and processing techniques can make feasible a wide range of analytics workloads already on one commodity machine. There is still much to explore in the case of one machine before turning to compute farms. A key practical benefit of this line of work is energy-efficient, inexpensive analytics over large databases.

The organization of the tutorial follows the structure of the next sections.

\section{Overview of Main Approaches to Machine Learning over Relational Databases}

The approaches highlighted in this tutorial are classified depending on how tightly they integrate the data system, where the input data reside and the training dataset is constructed, and the machine learning library (statistical software package), which casts the model training problem as an optimization problem. 

\subsection{No Integration of Databases and Machine Learning}

By far the most common approach to learning over relational data is to use two distinct systems, that is, the data system for managing the training dataset and the ML library for model training. These two systems are thus distinct tools on the technology stack with {\em no integration} between the two. The data system first computes the training dataset as the result of a {\em feature extraction query} and exports it as one table commonly in CSV or binary format. The ML library then imports the training dataset in its own format and learns the desired model. 

For the first step, it is common to use open source database management systems, such as PostgreSQL or SparkSQL~\cite{Spark:NSDI:2012}, or query processing libraries, such as Python Pandas~\cite{pandas} and R dplyr~\cite{dplyr}. Common examples for ML libraries include scikit-learn~\cite{scikit2011}, R~\cite{R-project}, TensorFlow~\cite{tensorflow}, and MLlib~\cite{MLlib:JMLR:2016}.

One advantage is the delegation of concerns: Database systems are used to deal with data, whereas statistical packages are for learning models. Using this approach, one can learn virtually any model over any database. 

The key disadvantage is the non-trivial time spent on materializing, exporting, and importing the training dataset, which is commonly orders of magnitude larger than the input database. Even though the ML libraries are much less scalable than the data systems, in this approach they are thus expected to work on much larger inputs. 
Furthermore, these solutions inherit the limitations of both of their underlying systems, e.g., the maximum data frame size in R and the maximum number of columns in PostgreSQL are much less than typical database sizes and respectively
number of model features.

\subsection{Loose Integration of Databases and Machine Learning}

A second approach is based on a {\em loose integration} of the two systems, with code of the statistical package migrated inside the database system space. In this approach, each machine learning task is implemented as a distinct user-defined aggregate function (UDAF) inside the database system. For instance, there are distinct UDAFs for learning: logistic regression models, linear regression models, $k$-means, Principal Component Analysis, and so on. Each of these UDAFs are registered in the underlying database system and there is a keyword in the query language supported by the database system to invoke them.
The benefit is the direct interface between the two systems, with one single process running for both the construction of the training dataset and learning. The database system computes one table, which is the training dataset, and the learning task works directly on it. Prime example of this approach is MADLib~\cite{MADlib:2012} that extends PostgreSQL with a comprehensive library of machine learning UDAFs. The key advantage of this approach over the previous one is better runtime performance, since it does not need to export and import the (usually large) training dataset. Nevertheless, one has to explicitly write a UDAF for each new model and optimization method, essentially redoing the large implementation effort behind well-established statistical libraries. Approaches discussed in the next sections also suffer from this limitation, yet some contribute novel learning algorithms that can be asymptotically faster than existing off-the-shelf ones.

A variation of the second approach provides a {\em unified programming architecture}, one framework for many machine learning tasks instead of one distinct UDAF per task, with possible code reuse across UDAFs. Prime example of this approach is Bismark~\cite{Kumar:InDBMS:2012}, a system that supports incremental (stochastic) gradient descent for convex programming. Its drawback is that its code may be less efficient than the specialized UDAFs. Code reuse across various models and optimization problems may however speed up the development of new functionalities such as new models and optimization algorithms.

\subsection{Tight Integration of Databases and Machine Learning}

The aforementioned approaches do not exploit the structure of the data residing in the database. The next and final approach features a {\em tight integration} of the data and learning systems. The UDAF for the machine learning task is pushed into the feature extraction query and one single evaluation plan is created to compute both of them. This approach enables database optimizations such as pushing parts of the UDAFs past the joins of the feature extraction query. Prime examples are Orion~\cite{KuNaPa15},which supports generalized linear models, Hamlet~\cite{Kumar:SIGMOD:16}, which supports logistic regression and na\"ive Bayes, Morpheus~\cite{Kumar:PVLDB:2017}, which linear and logistic regression, k-means clustering, and Gaussian non-negative matrix factorization, F~\cite{SOC:SIGMOD:16,OS:PVLDB:2016,OS:SIGREC:2016}, which supports ridge linear regression, AC/DC~\cite{ANNOS:DEEM:18}, which supports polynomial regression and factorization machines~\cite{RendleFM,libfm,Rendle13}, and LMFAO~\cite{lmfao}, which supports a larger class of models including the previously mentioned ones and decision trees~\cite{cart84}, Chow-Liu trees~\cite{Chow-Liu-trees:1968}, mutual information, and data cubes~\cite{DataCube:ICDE:1996,DataCube:SIGMOD:1996}. 

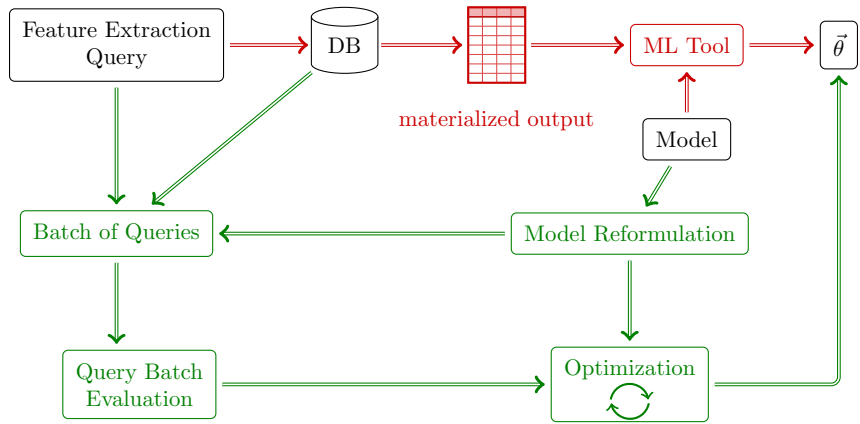
\begin{figure}[t]
\centering
\begin{tikzpicture}[yscale = 0.5, xscale=0.5, every node/.style={transform shape}]
  \node[data,scale=1.7] (FEQ) {Feature Extraction \\ Query};
  
  \begin{scope}[shift={($(FEQ)+(6,0)$)},scale=1,every node/.style={transform shape}]
    \node[inner sep=.3cm,scale=1.7] at (0,0) (DB) {DB};
    \drawcylinder{1.75}{1.25}{.25}{black}
  \end{scope}
  
  \begin{scope}[shift={($(DB)+(4,0)$)}, scale=1, every node/.style={transform shape}]
    \node[inner sep=.9cm] (0,0) (output_table){};
    \drawtable{1.5}{2}{4}{8}{clr_outofDB}
    \node[clr_outofDB, align=center, scale=1.7] at (0,-2) {materialized output};
  \end{scope}
  
  \node[data, clr_outofDB, scale=1.7] at($(output_table)+(5,0)$) (ML) {ML Tool};
  \node[data, scale=1.7] at($(ML)+(4,0)$) (theta_opt) {$\vec\theta$};
  
  \node[data, scale=1.7] at($(ML)+(0,-2.5)$) (model) {Model};

  \node[data,clr_inDB,scale=1.7] at($(model)+(-1.5,-2.5)$) (model_reform)
  {Model Reformulation};
  
  \node[data,clr_inDB,scale=1.7] at($(FEQ)+(0,-5)$) (FAQ)
  {Batch of Queries};
  
  \node[data, clr_inDB,scale=1.7] at($(FAQ)+(0.6,-4)$) (sigma_c)
  {Query Batch\\ Evaluation};
  
  \node[data, clr_inDB, scale=1.7] at($(model_reform)+(0,-4)$) (GD)
  {Optimization\\\vspace{.1cm}};
  \begin{scope}[shift={($(GD)+(0,-.45)$)},scale=.5,every node/.style={transform shape}]
    \draw[clr_inDB,->,thick] (-1,.1) arc (170:10:1);
    \draw[clr_inDB,->,thick] (1,-.1) arc (-10:-170:1);
  \end{scope}

  \draw[path, clr_outofDB] (FEQ)--(DB);
  \draw[path, clr_outofDB] (DB)--(output_table);
  \draw[path, clr_outofDB] (output_table)--(ML);
  \draw[path, clr_outofDB] (ML)--(theta_opt);
  \draw[path, clr_outofDB] (model)--(ML);

  \draw[path, clr_inDB] (model)--(model_reform);
  \draw[path, clr_inDB, scale=1.7] (model_reform)--(FAQ); 
  \draw[path, clr_inDB] (DB)--(FAQ);
  \draw[path, clr_inDB] (FEQ)--(FAQ);
  \draw[path, clr_inDB] (FAQ)--($(sigma_c.north)-(0.6,0)$);
  \draw[path, clr_inDB] (sigma_c)--(GD);
  \draw[path, clr_inDB, scale=1.7] (model_reform)--(GD); 
  \draw[path, clr_inDB] (GD) -| (theta_opt);
\end{tikzpicture}
\caption{Structure-aware versus structure-agnostic learning over relational databases.}
\label{fig:structure-aware}
\end{figure}

\section{Structure-aware Learning}

The tightly-integrated systems F~\cite{SOC:SIGMOD:16}, AC/DC~\cite{ANNOS:DEEM:18}, and LMFAO~\cite{lmfao} are {\em data structure-aware} in that they exploit the structure and sparsity of the database to lower the complexity and drastically improve the runtime performance of the learning process. In contrast, we call all the other systems {\em structure-agnostic}, since they do not exploit properties of the input database.
Figure~\ref{fig:structure-aware} depicts the difference between structure-aware (in green) and structure-agnostic (in red) approaches. The structure-aware systems compile the model specification into a set of aggregates, one per feature or feature interaction. This is called model reformulation in the figure. Data dependencies such as functional dependencies can be used to reparameterize the model, so a model over a smaller set of functionally determining features is learned instead and then mapped back to the original model. Join dependencies, such as those prevalent in feature extraction queries that put together several input tables, are exploited to avoid redundancy in the representation of join results and push the model aggregates past joins. The model aggregates over the feature extraction query define a batch of queries. In practice,
for training datasets with tens of features, query batch sizes can be in the order of: hundreds to thousands for ridge linear regression; thousands for computing a decision tree node; and tens for an assignment step in $k$-means clustering~\cite{lmfao}. The result of a query batch is then the input to an optimizer such as a gradient descent method that iterates until the model parameters converge.

Structure-aware methods have been developed (or are being developed) for a variety of models~\cite{ANNOS:PODS:2018}. Besides those mentioned above, powerful models that can be supported are: Principal Component Analysis (PCA)~\cite{murphy2013}, Support Vector Machines (SVM)~\cite{Joachims:2006}, Sum Product Networks (SPN)~\cite{SPN2011}, random forests, boosting regression trees, and AdaBoost. Newer methods also look at linear algebra programs where matrices admit a database interpretation such as the results of queries over relations. In particular, on-going work~\cite{Bas:thesis:2018,Gabriel:thesis:2019} tackles various matrix decompositions, such as QR, Cholesky, SVD~\cite{matrix2016comp}, and low-rank~\cite{glrm}.

Structure-aware methods call for new data processing techniques to deal with large 
query batches. Recent work puts forward new optimization and evaluation strategies that go beyond the capabilities of existing database management systems.
Recent experiments confirm this observation: Whereas existing query processing techniques are mature at executing one query, they miss opportunities for systematically sharing computation across several queries in a batch~\cite{lmfao}. 

Tightly-integrated DB-ML systems  commonly exploit four types of structure:  algebraic, combinatorial, statistical, and geometric.

\paragraph{Algebraic Structure}
The algebraic structure of semi-rings underlies the recent work on factorized databases~\cite{OlZa15,OS:SIGREC:2016}. The distributivity law in particular allows to factor out data blocks common to several tuples, represent them once and compute over them once. Using factorization, relations can represented more succinctly as directed acyclic graphs. For instance, the natural join of two relations is a union of Cartesian products. Instead of representing such a Cartesian product of two relation parts explicitly as done by relational database systems, we can represent it symbolically as a tree whose root is the Cartesian product symbol and has as children the two relation parts.
It has been shown that factorization can improve the performance of joins~\cite{OlZa15}, aggregates~\cite{BKOZ13,faq}, and more recently machine learning~\cite{SOC:SIGMOD:16,OS:SIGREC:2016,ANNOS:PODS:2018,faqai}. The additive inverse of rings allows to treat uniformly data updates (inserts and deletes) and enables incremental maintenance of models learned over relational data~\cite{Koch:DBToaster:2014,Nikolic:FIVM:2018,Kara:ICDT:2019}. The sum-product abstraction in (semi) rings allows to use the same processing (computing and maintaining) mechanism for seemingly disparate tasks, such as database queries, covariance matrices, inference in probabilistic graphical models, and matrix chain multiplication~\cite{faq,Nikolic:FIVM:2018}. The efficient maintenance of covariance matrices is a prerequisite for the availability of fresh models under data changes~\cite{Nikolic:FIVM:2018}. A recent tutorial overviews advances in incremental view maintenance~\cite{Iman:CIKM:2019}.

\paragraph{Combinatorial Structure}
The combinatorial structure prevalent in relational data has been formalized by notions such as width and data degree measures. If a feature extraction query has width $w$, then its data complexity is $\tilde{O}(N^w)$ for a database of size $N$, where $\tilde{O}$ hides logarithmic factors in $N$. Various width measures have been proposed recently, such as: the fractional edge cover number~\cite{GM06,AGM08,NPRR12,skew,LFTJ} to capture the asymptotic size of the results for join queries and the time to compute them; the fractional hypertree width~\cite{Marx:FHTW:2010} and the submodular width~\cite{panda17} to capture the time to compute Boolean conjunctive queries; the factorization width~\cite{OlZa15} to capture the size of the factorized results of conjunctive queries; 
the FAQ-width~\cite{faq} that extends the factorization width from conjunctive queries to functional aggregate queries; and the sharp-submodular width~\cite{faqai} that improves on the previous widths for functional aggregate queries.
 
The degree information captures the number of occurrences of a data value in the input database~\cite{skew}. Existing processing techniques adapt depending on the high or low degree of data values. A recent such technique has been shown to be worst-case optimal for incrementally maintaining the count of triangles in a graph~\cite{Kara:ICDT:2019}. Another such technique achieves a low complexity for computing queries with negated relations of bounded degree~\cite{Khamis:ICDT:2019}. A special form of bounded degree is given by functional dependencies, which can be used to reparameterize (polynomial regression and factorization machine) models and learn simpler, equivalent models instead~\cite{ANNOS:PODS:2018}.

\paragraph{Statistical Structure}
The statistical structure allows to sample through joins, such as the ripple joins~\cite{Haas:RippleJoin:SIGMOD:1999} and the wander joins~\cite{Li:WanderJoin:TODS:2019}, and to sample for specific classes of machine learning models~\cite{Park:SampleML:SIGMOD:2019}. 
Sampling is employed whenever the input database is too large to be processed within a given time budget. It may nevertheless lead to approximation of both steps in the end-to-end learning task, from the computation of the feature extraction query to the subsequent optimization task that yields the desired model.
Work in this space quantifies the loss in accuracy of the obtained model due to sampling.

\paragraph{Geometric Structure} 
Algorithms for clustering methods such as $k$-means~\cite{murphy2013} can exploit distance measures (such as the optimal transport distance between two probability measures) to obtain constant-factor approximations for the $k$-means objective by clustering over a small grid coreset instead of the full result of the feature extraction query~\cite{Curtin:2019}.

\section{Database Systems Considerations}

Besides exploiting the structure of the input data and the learning task, the problem of learning models over databa\-ses can also benefit tremendously from database system techniques. 
Recent work~\cite{lmfao} showed non-trivial speedups (several orders of magnitude) brought by code optimization for machine learning workloads over state-of-the-art systems such as TensorFlow~\cite{tensorflow}, R~\cite{R-project}, Scikit-learn~\cite{scikit2011}, and mlpack~\cite{mlpack2018}. Prime examples of code optimizations leading to such performance improvements include:

\paragraph{Code Specialization and Query Compilation}  
It involves generating code specific to the query and the schema of its input data, following prior work~\cite{Neumann:PVLDB:11,legobase_tods,dblablb}, and also specific to the model to be learned.
This technique improves the runtime performance by inlining code and improving cache locality for the hot data path.

\paragraph{Sharing Computation} 
Sharing is best achieved by decomposing the aggregates in a query batch into simple views that are pushed down the join tree of the feature extraction query. Different aggregates may then need the same simple views at some nodes in the join tree. Sharing of scans of the input relations can also happen across views, even when they have different output schemas.

\paragraph{Parallelization} 
Parallelization can exploit multi-core CPU architectures but also large share-nothing distributed systems. It comprises both task parallelism, which identifies subqueries that are independent and can be computed in parallel, and domain parallelism, which partitions relations and computes the same subqueries over different parts in parallel.

\paragraph{This tutorial is a call to arms for more sustained and principled work on the theory and systems of structure-aware approaches to data analytics. What are the theoretical limits of structure-aware learning? What are the classes of machine learning models that can benefit from structure-aware learning over relational data? What other types of structure can benefit learning over relational data?}

%
%
%
\bibliographystyle{plain}
%

\end{document}